\documentclass[aps,superscriptaddress]{revtex4-1}

\usepackage{amsbsy}
\usepackage{amsfonts}
\usepackage{amssymb}
\usepackage{amsmath}
\usepackage{color}
\usepackage{graphicx}
\usepackage{xspace}
\usepackage[normalem]{ulem}

\newcommand{\ket}[1]{| #1 \rangle}
\newcommand{\bra}[1]{\langle #1 |}

\begin{document}

\title{Reversible work extraction in a hybrid opto-mechanical system}

\author{Cyril Elouard}
\affiliation{CNRS and Universit\'e Grenoble Alpes, Institut N\'eel, F-38042 Grenoble, France}

\author{Maxime Richard}
\affiliation{CNRS and Universit\'e Grenoble Alpes, Institut N\'eel, F-38042 Grenoble, France}

\author{Alexia Auff\`eves}
\email{alexia.auffeves@neel.cnrs.fr}
\affiliation{CNRS and Universit\'e Grenoble Alpes, Institut N\'eel, F-38042 Grenoble, France}

\begin{abstract}
With the progress of nano-technology, thermodynamics also has to be scaled down, calling for specific protocols to extract and measure work. Usually, such protocols involve the action of an external, classical field (the battery) of infinite energy,  that controls the energy levels of a small quantum system (the calorific fluid). Here we suggest a realistic device to reversibly extract work in a battery of finite energy : a hybrid optomechanical system.  Such devices consist in an optically active two-level quantum system interacting strongly with a nano-mechanical oscillator that provides and stores mechanical work, playing the role of the battery. We identify protocols where the battery exchanges large, measurable amounts of work with the quantum emitter without getting entangled with it. When the quantum emitter is coupled to a thermal bath, we show that thermodynamic reversibility is attainable with state-of-the-art devices, paving the road towards the realization of a full cycle of information-to-energy conversion at the single bit level.
\end{abstract}

\maketitle

\section{Introduction}

\begin{figure}[t]
\begin{center}
\includegraphics[width=0.5\columnwidth,natwidth=1298,natheight=505]{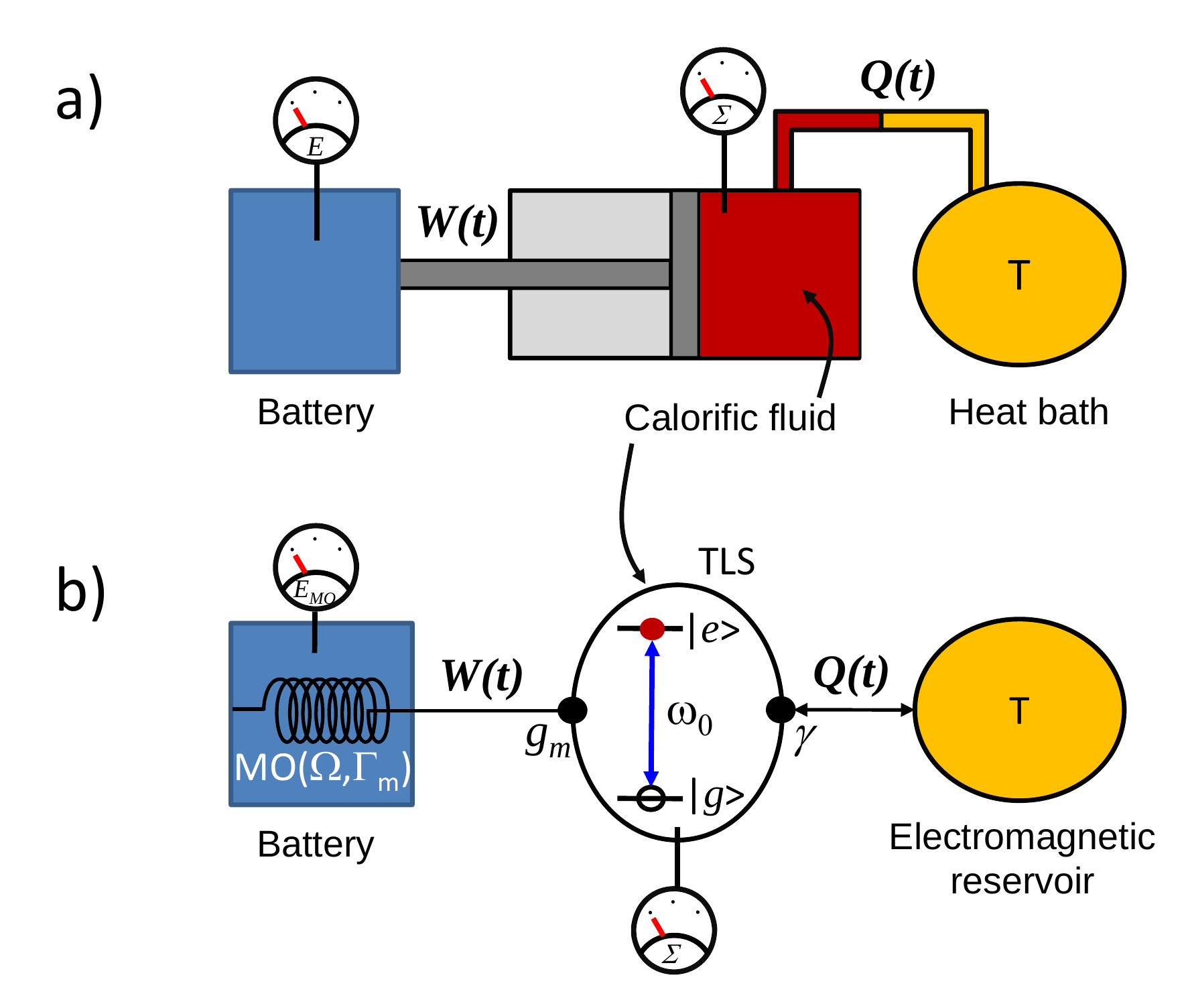}
\end{center}
\caption{a) A typical experimental setup in thermodynamics. A system (the calorific fluid) exchanges heat $Q$ with a bath, and work $W$ with a battery. In the framework of information thermodynamics, the system has two micro-states allowing to encode one bit of information. Mean work exchanges are usually studied by recording and processing the mean trajectory $\Sigma(t)$ of the system all along the transformation. In our protocol, work exchanges are inferred from the readout of the battery states at the initial time $t_i$ and final time $t_f$. b) A hybrid opto-mechanical system : an optically active two-level system (TLS) of transition frequency $\nu_0$ is coupled to a mechanical oscillator (MO) of frequency $\Omega$ with a strength $g_m$. The TLS interacts with a reservoir of electromagnetic modes, its spontaneous emission rate being denoted $\gamma$. In the protocol that we suggest, information is encoded on the TLS, while the MO (resp. the electromagnetic reservoir) plays the role of the battery (resp. of the bath).}
\label{fig0}\end{figure}

Thermodynamics was born in the $19^\mathrm{th}$ century, with the practical purpose of understanding the mechanism governing the conversion of heat present in reservoirs of disorganized energy, into useful mechanical work extracted in reservoirs of organized energy, by exploiting the transformations of a calorific fluid (Fig.\ref{fig0}a). This initially applied area of physics, aimed at building engines, was later shown to have a deeper content. Time arrow and the concept of irreversibility are such byproducts. In the $20^\mathrm{th}$ century, Szilard \cite{Szilard} and Landauer \cite{Landauer} have shown in their pioneering works that calorific fluids could be identified with microscopic systems encoding bits of information. As a result, they found out that one bit of information could be reversibly converted into an elementary amount of energy. The validity of this principle has been recently extended to information of quantum nature, such that coherence and entanglement are also expected to have energetic counterparts \cite{DelRio, Oppenheim, Zurek}. These theoretical results form the core of quantum information thermodynamics.

This blooming field of quantum (information) thermodynamics calls for the development of dedicated experimental platforms, involving a single quantum system as a calorific fluid, interacting with one or several heat baths and batteries. Nowadays, an increasing number of experimental setups realize this situation, such as superconducting Cooper pair boxes \cite{Koski}, ion traps \cite{Abah}, or nuclear magnetic resonance setups \cite{Celeri}. In such experiments, the interesting thermodynamic quantities are the average fluxes of work and heat, as well as their probability distributions. Intense efforts have been carried out in order to determine how such quantities should be measured \cite{Talkner, Solinas-PRB, Suomela, Frenzel, Binder}. In pioneering experiments performed in the classical regime, heat and work exchanges are deduced from time-resolved measurements of the bit state throughout the whole thermodynamic transformation. Work is provided by an external operator, i.e. a battery of quasi-infinite energy \cite{Berut, Ueda, Koski}, that undergoes no back-action from the system. This strategy allows full control of the transformation timescale, such that reversibility is reached. In the quantum world, such measurements are especially challenging, since the observation of quantum trajectories requires to readout the qubit states in a quantum non-demolition manner on ultra-fast timescales \cite{Haroche,Siddiqi}. More recently, alternative strategies have been suggested \cite{Dorner, Mazzola} and implemented \cite{Celeri, Batalhao}, based on coherent control of an ancillary quantum system to access work or heat distributions.

In this article, we investigate another strategy to measure average work exchange, that is based on reversible work extraction in a battery of finite energy, undergoing noticeable back-action from the system. Instead of deriving work from the time-resolved measurement of the system state, or the readout of an ancillary qubit state, we show that only two measurements of the battery's energy, at the initial and final times of the thermodynamic transformation, are required. With this aim, we show that a hybrid opto-mechanical device constitutes a very suitable experimental platform. Such a system consist in an optically active quantum emitter carrying one bit of information, coupled to a nano-mechanical oscillator playing the role of the battery. The emitter also interacts with the electromagnetic (EM) field prepared in a thermal state, that plays the role of the heat bath. Owing to its small size and its large interaction strength with the quantum system, we expect the battery to be visibly affected by the elementary work produced during the conversion of a single bit of information. Since it involves a quantum bit, such an hybrid system could also be used for the experimental exploration of quantum information to energy conversion.

The paper is organized as follows: first, we do not consider the heat bath and we solely characterize the energy transfer between the nanomechanical oscillator and the quantum system (adiabatic transformation). Then, we couple the emitter to a thermal heat bath of finite temperature T and study its isothermal transformations. We show that the periodic dynamics of the device can be seen as series of Landauer's erasures and Szilard engines, where one bit of information is reversibly converted to an elementary work stored in/extracted from the mechanical oscillator.

\section{Adiabatic transformations}

\subsection{System and model}
A hybrid optomechanical device features a quantum two-level system (TLS) whose ground and excited levels are respectively denoted $\ket{g}$ and $\ket {e}$, and frequency transition $\nu_0$. The TLS is coupled to a mechanical oscillator (MO) of frequency $\Omega$, whose mean deflection is further denoted $x(t)$. The oscillator zero point fluctuation is noted $x_0$. Actual realizations of such devices are based on semiconductor quantum dots \cite{Yeo} or diamond Nitrogen-Vacancies (NV) \cite{Arcizet} coupled to a vibrating wire, or superconducting qubits embedded in oscillating membranes \cite{Sillanpaa}.

The total Hamiltonian of the system can be written as
\begin{align}
H_0= H_m + H_q + V,
\end{align}
where $H_q = \frac {h \nu_0}{2} (\sigma_z + 1)$ is the free Hamiltonian of the TLS,  $V= \frac{h g_m}{2} (\sigma_z + 1) (b^\dagger + b)$ is the TLS-mechanical coupling according to the spin-boson Hamiltonian, where $g_m$ is the coupling strength, and $H_m = h \Omega (b^\dagger b + \frac{1}{2})$ is the free Hamiltonian of the MO. We have introduced the spin one-half operator $\sigma_z = \ket{e}\bra{e} - \ket{g}\bra{g}$. $b$ is the lowering (phonon annihilation) operator in the mechanical mode, such that $x_0(b+b^\dagger) = \hat{x}$, where $\hat{x}$ is the MO deflection operator.

We are interested in the regime where $\nu_0 \gg g_m \geq \Omega$. This is the so-called dispersive, ultra-strong coupling regime, which is within reach in state-of-the-art devices \cite{Yeo, Sillanpaa}. To solve the dynamics of the coupled system, it is useful to notice that this situation is highly reminiscent of the dynamics of electrons in vibrating molecules studied by Born and Oppenheimer \cite{Messiah}. In this textbook case, adiabatic theorem applies, stating that electrons evolve in the time-dependent potential defined by the mean position of the nuclei, while the nuclei's motion depends on the electronic position averaged over several trajectories. Similarly here, $H_m$ is a first-order perturbation of the Hamiltonian $H_q+V$ of eigenstates $\ket{g,x}$ and $\ket{e,x}$ with respective eigenvalues $0$ and $h(\nu_0+ g_m x/x_0)$. The perturbation $H_m$ couples states of different $x$, while $\ket{g}$ and $\ket{e}$ remain "good quantum states", defining two subspaces ${\cal H}_g$ and ${\cal H}_e$ stable under the total dynamics of the coupled system. It is therefore possible to split the MO and the TLS dynamics, by using the polaron transformation.

The MO dynamics is governed by the effective Hamiltonian $H^g_m = h \Omega (b^\dagger b+\frac{1}{2})$ if the TLS is in $\ket{g}$ or $H^e_m =h \Omega \left(B^\dagger B+ \frac{1}{2} - (g_m/\Omega)^2\right)$ if the TLS is in $\ket{e}$. The displaced lowering operator is defined as $B = b + g_m/\Omega$; it captures the fact that when the TLS is in state $\ket{e}$ it generates a static field which shifts the rest position of the MO by $-2 x_0g_m/\Omega$. Note that in this regime of coupling, if the MO has been initially prepared in a coherent or in a thermal state, the TLS does not change its classical nature \cite{qhammer}. On the other hand, the TLS dynamics is governed by a time-dependent Hamiltonian
\begin{align}
H_q(t) = H_q + \mathrm{Tr}_m(\rho_m(t) V) = \left( \frac{h \nu_0}{2}  + \frac{h g_m x(t)}{2 x_0} \right) (\sigma_z + 1),
\end{align}
where we have introduced the reduced density matrix of the MO $\rho_m(t)=\mathrm{Tr}_q[\rho(t)]$, where $\rho(t)$ is the total density matrix of the coupled system. Therefore the main effect of the MO on the TLS is to modulate its transition energy by an amount $\delta(t) = \frac{h g_m x(t)}{x_0}$.

\subsection{Average work exchanges}
From the study above, it appears that the motion of the MO results in a time-dependent Hamiltonian acting on the TLS. Introducing the reduced density matrice of the TLS $\rho_q(t)=\mathrm{Tr}_m[\rho(t)]$, and according to the usual definition of average work $\dot{w} = Tr_q[\rho_q(t) \dot{H}_q(t)]$ \cite{Solinas-PRB,Frenzel, Binder}, the MO provides work to the TLS with the rate $\dot{w} = (h g_m \dot{x}/x_0) P_e$, where $P_e(t) = \mathrm{Tr}_q(\rho_q(t)(\sigma_z+1)/2)$ is the excited emitter population. In this part, we do not consider any heat exchanges with external thermal baths yet, and focus on the case of adiabatic transformations (in the thermodynamic sense): in this situation, the adiabatic work $W_{\mathrm{ad}}$ received by the TLS is entirely stored in the TLS internal energy, resulting in a transition frequency shift $\Delta \nu_0 = (g_m \Delta x/x_0) P_e$ such that $W_{\mathrm{ad}} = h\Delta \nu_0$.

We now provide another operational interpretation of the adiabatic work, based on the mechanical energy variation: let us consider a thermodynamic transformation where the TLS is initially excited, while the MO is prepared in a large coherent field corresponding to an initial elongation $x_m$ (see Fig.\ref{fig2}). As explained above, the TLS remains in $\ket{e}$ while the MO oscillates at its eigenfrequency $\Omega$ around the displaced equilibrium position $-2x_0g_m/\Omega$. Since the rest position is shifted, the MO oscillates between the position $x_m$ and $x_M = (-4 g_m/\Omega)x_0 - x_m$, which correspond respectively to a minimum and a maximum of the TLS transition frequency, and to the initial and final times $t_i=0$ and $t_f=1/(2\Omega)$ of the transformation.

During this motion, the mean energy of the total system is conserved: $\dot{\langle H_0 \rangle}  = 0 = \mathrm{Tr}[\dot{\rho}(H_q+H_m+V)]$, such that $0 = \mathrm{Tr}_m[\dot{\rho}_m (H_m + \mathrm{Tr}_q (\rho_q V)] + \mathrm{Tr}_q[\dot{\rho}_q (H_q + \mathrm{Tr}_m (\rho_m V)]$. Since $\dot{\rho}_q = 0$, we have finally $\mathrm{Tr}_m[\dot{\rho}_m H_m] =  -\mathrm{Tr}_m[ \dot{\rho}_m \mathrm{Tr}_q (\rho_q V)] $. After integration, the first term is identified with the variation of the average mechanical energy $\Delta E_m = E_m(t_f)-E_m(t_i)$, while the second term equals $hg_m \Delta x/x_0
$, which exactly corresponds to the work received by the TLS. Therefore we have $W_{\mathrm{ad}} = - \Delta E_m$, clearly showing that in this thermodynamic transformation, the MO is the battery providing work to the TLS. As mentioned above, this quantity also equals the TLS energy variation, and can thus be written $W_{\mathrm{ad}}= h \Delta \nu_0 = 2g_m(2g_m/\Omega + x_m/x_0)$. This simple study evidences that a hybrid device can be used as a ultimate opto-mechanical transducer, able to convert optical quanta of energy into mechanical ones. This conversion is reversible : during the extension step, work is extracted from the mechanics into the TLS, while in the compression step, work is extracted from the TLS and stored in the mechanics.	

\subsection{Otto cycle}

\begin{figure}[t]
\begin{center}
\includegraphics[width=0.35\columnwidth,natwidth=1298,natheight=505]{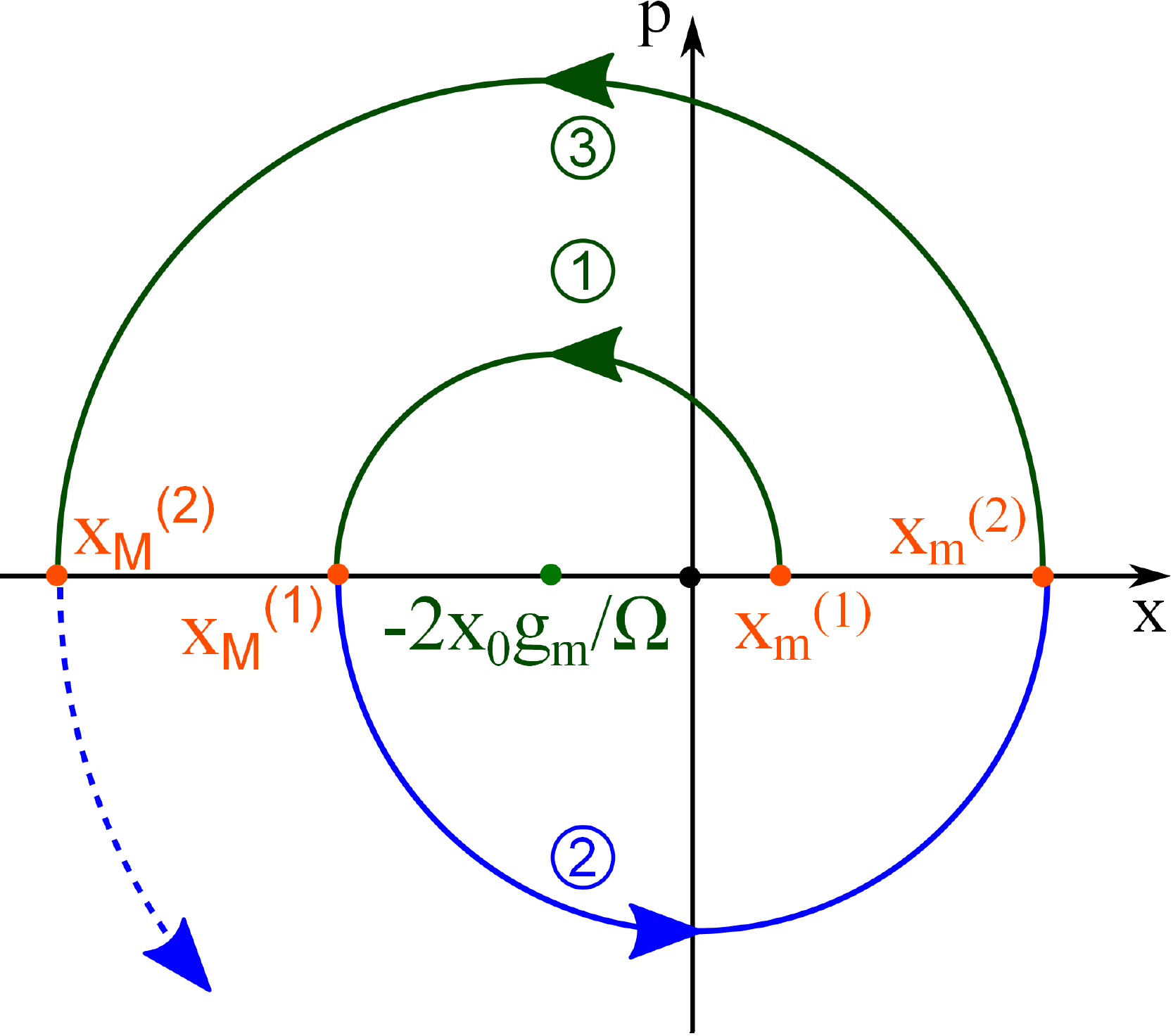}
\end{center}
\caption{An Otto cycle with a hybrid optomechanical system. When the TLS is in the excited state, the MO rotates around a displaced rest position (green lines). When the TLS is in the ground state, the MO rotates around the initial rest position (blue lines). The amplitude of the mechanical oscillation increases at each iteration.}
\label{fig2}\end{figure}

The mechanism studied above can be exploited to realize an Otto cycle (see Fig.\ref{fig2}) using the following protocol:  at $t=t_1$, the MO is shifted by $x^{(1)}_m$ from its rest position $x=0$. The TLS is heated up and prepared in a fully mixed state $\rho_q(t_1) = \mathbb{I}/2$. Here we suppose that heating and cooling of the TLS can be performed on timescales much shorter  than the mechanical period, which is usually realized in opto-mechanical systems. Then the bath is decoupled and the adiabatic transformation starts. If the TLS is in the ground state, no work is exchanged. If the TLS is in the excited state, the MO oscillates around its shifted rest position until maximal elongation $x^{(1)}_M$ is reached at time $t_1+1/(2\Omega)$, extracting a work $W_{\mathrm{ad}}$. On average, a mean work $W_1 = W_{\mathrm{ad}}/2 = h g_m(2g_m/ \Omega+x^{(1)}_m/x_0)$ is extracted from the TLS. At this moment, the TLS is cooled down to the ground state, for instance by tuning a cavity on resonance with the TLS to enhance spontaneous emission by Purcell effect. The MO keeps oscillating at the frequency $\Omega$, now around its initial rest position $x=0$, until reaching $x_m^{(2)}=2x_0g_m/\Omega + x^{(1)}_m$ where the TLS is heated up again, allowing to extract another amount of work $W_2$. Hence by modulating the emitter population inversion, optical excitation of the mechanics can be reached : a similar effect was evidenced in \cite{qhammer}, in the regime of small TLS-mechanical coupling. The net work extracted from the engine increases at each iteration, and equals after $n$ iterations $W_n = hg_m(2n g_m/\Omega + x^{(1)}_m/x_0)$. Note that we neglect the mechanical relaxation, which is valid as long as the number of oscillations remains low with respect to the mechanical quality factor $Q_m\sim 1000$. The typical power of the engine scales like $P = W\Omega$. Using realistic parameters \cite{Yeo}, we find that in $100$ iterations, and starting from the rest position of the MO, the typical power increases from $P \sim 10^{-21} W$ to $10^{-19}$W, a value which is consistent with other proposals involving ion traps \cite{Abah}.

\section{Isothermal transformations}

In the former section, we focused on adiabatic transformations and evidenced that the MO can be seen as a battery, as work provided to the TLS corresponds to average mechanical energy variations between the initial time and final time of the transformation. Here we extend the study to the case of isothermal transformations, and couple the TLS to a heat bath of temperature $T$.

\subsection{Master equations}

The total Hamiltonian for the hybrid opto-mechanical system and the bath is

\begin{align}
H = H_0+H_{q-R} + H_R
\end{align}	

where $H_0$ is the Hamiltonian of the opto-mechanical system as defined above, $H_{q-R}$ is the coupling Hamiltonian between the TLS and the heat bath, and $H_R$ is the Hamiltonian of the bath. In the absence of TLS-mechanical coupling ($g_m = 0$), the effect of the bath on the emitter would be described by the Lindblad super-operator ${\cal L}[\rho_q] = \gamma (\bar n+1)D[\sigma]\rho + \gamma \bar n D[\sigma^\dagger]\rho$, where we have introduced the super-operator $D[X]\rho = X\rho X^\dagger - \tfrac 1 2 (X^\dagger X\rho + \rho X^\dagger X)$. $\gamma$ is the spontaneous emission rate of the TLS and satisfies $\gamma\ll \nu_0$. $\bar{n}$ is the average number of thermal excitations in the EM reservoir in the mode resonant with the TLS frequency. As $|g_m|>0$, a new master equation has to be derived to take into account the modifications induced by the MO on the TLS dynamics \cite{Solinas-PRL, Alicki, Levy}. In what follows, we are interested in the case where $\gamma\gg g_m$. This situation corresponds to the semi-classical regime : the bath measures the TLS in the $\ket{e},\ket{g}$ basis, on a typical timescale $\gamma^{-1}$. This is to be compared with the characteristic time $g_m^{-1}$ for correlations to build up between the MO and the TLS. Since $g_m^{-1}\gg \gamma^{-1}$, these correlations can be safely neglected. Following \cite{Wallquist,Rabl}, we therefore use a mean field approach, and approximate the state of the coupled MO-TLS-bath system by projecting it on the factorized density matrix $\rho_m\otimes\rho_q\otimes \rho_R$, where $\rho_R$ is the density matrix of the EM bath. Injecting this Ansatz in the dynamical equation $\dot{\rho} = -i[H,\rho]$, we first decouple the MO from the TLS-bath evolution and write its dynamic as:

\begin{align}
\dfrac d {dt} \rho_m(t) =& -i[H_m + \mathrm{Tr}_q(\rho_q V),\rho_m] = -i[\Omega b^\dagger b + g_m P_e(t) (b+b^\dagger),\rho_m]. \label{eq:Mm}
\end{align}

We then write the TLS-bath dynamics
\begin{align}
\dfrac d {dt}\left( {\rho_q\otimes\rho_R}\right) =&  -i[ H_q + \mathrm{Tr}_m(\rho_m V) +H_{q-R} + H_R , \rho_q \otimes \rho_R] = -i\left[ \frac{h(\nu_0+\delta(t))}{2} (\sigma_z+1)  +H_{q-R}+ H_R  , \rho_q \otimes \rho_R\right] \label{eq:Mq}
\end{align}

where the mean TLS population $P_e(t)$ and TLS frequency shift $\delta(t)$ are:

\begin{align}
P_e(t) = \mathrm{Tr}_q\left[\frac{\sigma_z+1}{2} \rho_q(t)\right],
\end{align}
and
\begin{align}
\delta(t) = g_m \mathrm{Tr}_m\left[(b+b^\dagger)\rho_m(t)\right].
\end{align}

In agreement with the previous section, Eq.(\ref{eq:Mm}) shows that the TLS generates an optical force on the MO, that displaces it from its rest position. The pump term proportional to $P_e(t)$ is analogous to the radiation pressure in the context of cavity opto-mechanics \cite{Rabl,qhammer}. Note that by pushing the development to the second order, it is possible to study the fluctuations of the MO induced by the fluctuations of the quantum emitter, but this is beyond the scope of this paper. Here, our goal is to calculate the average energy exchanges, we therefore only keep track of the first order.

Again in agreement with the first section, Eq.\eqref{eq:Mq} shows that the MO modulates the TLS transition frequency by an amount $\delta(t)$, inducing an effective time-dependent Hamiltonian $H_q(t) = \frac{h(\nu_0+\delta(t))}{2}(\sigma_z+1)$ that acts on the TLS. This effective modulation is a strong signature of the TLS-mechanical coupling on the optical properties of the quantum emitter. It has been observed both in the case of a semi-conductor quantum dot \cite{Yeo}, and of a NV center \cite{Arcizet}. In practice, observing such modulation requires a large coupling strength $g_m$, such that $|\delta| \gg \gamma$. Note that this condition is perfectly consistent with the semi-classical regime, defined as $g_m\ll \gamma$. Our study provides therefore an extension of \cite{Wallquist, Rabl} where this modulation was neglected.

The typical evolution timescale of $H_q(t)$ being the mechanical one, it largely overcomes the typical correlation time within the heat bath. Consequently, the bath can be reasonably traced out, resulting in the following master equation for the TLS:

\begin{align}
\dot{\rho}_q = - i[H_q(t),\rho_q] + {\cal L}_t[\rho_q] = - i[H_q(t),\rho_q] +  \gamma(t) (\bar n(t)+1)D[\sigma]\rho_q + \gamma(t) \bar n(t) D[\sigma^\dagger]\rho_q.\label{eq:Mq2}
\end{align}

As expected, the Lindbladian modeling the interaction with the bath is now time-dependent : the coupling parameters depend on $\gamma(t)$ and $\bar n(t)$, i.e. the spontaneous emission rate and average number of thermal excitations in the EM bath at the frequency $(\nu_0 + \delta(t))$. Finally we are able to derive time evolutions of operators expected values, from the effective master equations \eqref{eq:Mm} and \eqref{eq:Mq}:

\begin{align}
\dot{\beta} &= -i\Omega\beta -ig_m P_e   \label{eq:Bloch-beta}\\
\dot{N} &=  -ig_m P_e \left(\beta^* - \beta\right)\label{eq:Bloch-N}\\
\dot{P}_e &= -\gamma(t)(2\bar{n}(t) +1)   P_e + \gamma(t) \bar{n}(t)\label{eq:Bloch-sz}\\
\dot{s}  &=  - i(\delta(t)+\nu_0)s  - \frac{\gamma(t)}{2} (2\bar{n}(t) +1)s \label{eq:Bloch-s}.
\end{align}

We have introduced $s(t) = \mathrm{Tr}_q[\rho_q(t) \sigma]$ the mean TLS dipole amplitude where $\sigma = \ket{g}\bra{e}$ is the TLS dipole operator,  $\beta(t) = \mathrm{Tr}_m[\rho_m(t)  b]$ and $N(t) = \mathrm{Tr}_m [\rho_m(t) b^\dagger b ]$ the mean MO amplitude and population respectively.

\subsection{First principle of thermodynamics}
Owing to the above analysis, we can now evaluate the average energy exchanges between the TLS, the MO and the EM bath from a thermodynamic point of view. In general, the first principle applied on a quantum system of density matrix $\rho(t)$ driven by a time-dependent Hamiltonian $H(t)$ whose coupling to a reservoir is described by a Lindbladian ${\cal L}_t[\rho]$ can be written \cite{Solinas-PRB,Binder} $\dot{u} = \dot{w}+\dot{q}$,
where $u = \mathrm{Tr}[H(t)\rho(t)]$ is the internal energy of the TLS, $\dot{q} = \mathrm{Tr}{{\cal L}_t[\rho]H(t)}$ is the heat exchange rate with the reservoir, and $\dot{w}=\mathrm{Tr}{\rho\dot{H}}$ the work exchange rate between the TLS and the MO. Like in the first section, we have $\dot{w} = \mathrm{Tr}(\rho_q \dot{H_q}(t)) = h P_e(t) \dot{\delta}$. We consider an isothermal transformation defined by the evolution of the opto-mechanical device between the initial time $t_i=0$ and final time $t_f=t$. The total work exchange can be written $w(t) = \int_0^t du h \dot{\delta}(u) P_e(u)$. Such quantity can be measured by recording the trajectory of the TLS, i.e. its state all along the thermodynamical transformation, as it was for instance done in \cite{Berut,Koski}. However in our case, we can also measure the average mechanical energy $E_{m}(t)$. Taking into account Eq.\eqref{eq:Bloch-N}, the derivative $\dot{E}_{m}(t)$ reads $\dot{E}_{m}(t) =  - h\dot{\delta}(t) P_e(t)$. We thus find that $E_{m}(t) - E_{m}(0) =  -w(t)$. This extends the results of the previous section to the case of isothermal transformations : here again, the work performed on the TLS is fully provided by the MO, which therefore behaves as a proper battery. It clearly establishes that in principle, an integrated quantity like work, that depends on the full evolution of the system, can be directly read out by performing only two measurements, namely, the mechanical energy at the initial and final times of the transformation. In another context, this mechanical energy variation is identified as the back-action of the TLS on the mechanical motion. Such phenomenon has been extensively studied in the pioneering works of cavity opto-mechanics \cite{Kippenberg}. On the other hand, heat is emitted as photons of energy close to the TLS bare transition one, in the electromagnetic environment. In principle, such heat exchanges can also be recorded by accurate spectroscopic/calorimetric measurements \cite{Hekking}.

In the last following sections, we show that such isothermal transformations are expected to be reversible with state of the art opto-mechanical systems, and consider the possible applications in terms of information-to-energy conversions.

\subsection{Landauer erasure, Szilard engine : general principle}

Let us first remind the original Landauer erasure and Szilard engine protocols (see also ref. \cite{DelRio}). They involve a classical TLS of levels denoted 0 and 1. The Shannon entropy of the TLS is $H[p] = -p \log_2 (p )-(1-p) \log_2(1-p)$, where $\log_2$ is the logarithm in base 2, and $p$ the occupation probability of the state 1. In Landauer's protocol, both states are initially equiprobable, so that the initial entropy is $H_i = 1$. Then, the TLS is reset to a determined state, such that its final entropy is $H_f=0$. To do so, the TLS is coupled to an external battery at the initial $t_i = 0$, and put in contact with a thermal bath of temperature $T \gg E(0)/k_\mathrm{B}$, where $E(t)=E_1(t)-E_0$ is the energy difference at time $t$ between both levels and $k_\mathrm{B}$ the Boltzmann constant. $E$ is increased over time until it largely overcomes the thermal energy at the final time $t_f=t$. Then the TLS has fully relaxed into the state 0, so that initialization is complete. If the operation is performed slowly enough with respect to the thermalization time, thermodynamic equilibrium is always realized. Therefore the probability $P_1$ to find the system in state 1 follows a Fermi-Dirac distribution, reading $P_1^\infty (E) = e^{- E/k_\mathrm{B}T} /(1+e^{ - E/k_\mathrm{B} T} )$. During this operation, the battery provides a work $w(t) = \int_0^{t} dEP_1^\infty(E)$, which in this quasi-static situation reaches Landauer's limit $W_0 = k_\mathrm{B}T \log 2(H_i - H_f )$. Eventually, this work is dissipated as heat in the thermal bath $Q_0 = - W_0 = \int_0^t E dP_1^\infty (E)$. The reverse transformation is known as Szilard engine: the TLS is initially in a determined state, it is then coupled to a heat bath while its two levels are slowly brought back to degeneracy. In the end of this operation, the information on the initial state is lost and the TLS is in a mixed state 0 and 1, while an elementary work $W_0$ has been extracted from the bath and stored in the battery.

Landauer's erasure and Szilard engines are fundamental Gedanken experiments, where elementary amounts of information, quantified by the Shannon entropy of the encoding system, can be reversibly converted into elementary amounts of energy $W_0$. Later, their validity has been extended to quantum information, Shannon entropy being replaced by Von Neumann entropy. It is the basis of many protocols investigating the operational value of coherence and entanglement \cite{DelRio, Oppenheim, Zurek}. In these experiments, the proportionality between energy and entropy is crucial, and entirely depends on the reversibility of the process: if the transformations are performed too fast, energy is dissipated into irreversible mechanisms, such that proportionality is broken.

Despite their fundamental character, experimental evidences of reversible information-to-energy conversions in a battery of finite energy have remained elusive. If Landauer's minimum work has been measured \cite{Berut}, Szilard engines based on work extraction in a battery of finite size are still degraded by irreversible mechanisms, such that best information-to-work ratios nowadays reach $28\%$ \cite{Ueda}. More recently, optimal work extraction has been demonstrated by Koski et al \cite{Koski}, for a single electron coupled to a classical battery of infinite size. These results are fully consistent with the specific protocol studied in \cite{Frenzel}, where work extraction is theoretically shown to be optimal in the limit of a classical battery, i.e. undergoing no back-action from the system. As we show below, opto-mechanical devices also allow realizing optimal work extraction, i.e. reaching thermodynamic reversibility, while the battery's energy is finite.

\subsection{Monitoring reversible information-to-energy conversions}
In this section we focus on the evolution of the mechanical state over an entire mechanical period, and we show that a full cycle of information-to-energy conversions, corresponding to the situation described by Eq.\eqref{PPT}, can be evidenced. The electromagnetic field is in a thermal state of temperature $T$, such that the system is well described by equations \eqref{eq:Bloch-beta} to \eqref{eq:Bloch-s}. The MO is initially pulled out of its rest position. For $\Omega \ll \gamma$ (quasi-static condition), the TLS population follows the mechanical motion. Therefore the TLS is in thermal equilibrium with the heat bath at any time, and $P_e (t) = P_1^\infty (E (t))$, where $E(t) = h (\nu_0 + \delta(t))$ is the TLS transition energy. The MO oscillation modulates the TLS population between $P_1^\infty (E_{min})$ and $P_1^\infty (E_{max})$, where $E_{min}$ (resp. $E_{max})$ is the minimal (resp. maximal) energy of the TLS transition. The TLS population and entropy decrease when $\delta$ increases: this sequence corresponds to a Landauer erasure. During this step, work is extracted from the MO, leading to a decrease of the mechanical energy and elongation. Denoting again $x_m$ and $x_M$ the MO minimal and maximal elongations, the work extracted from the mechanics reads $W = -\Delta E_m = \frac{h\Omega}{4x_0^2}(x_M^2-x_m^2)$.
Conversely, the TLS population and entropy increase when $\delta$ decreases: this sequence corresponds to Szilard engine, and to an increase of the mechanical energy. If the transformations are reversible, work exchanges during Landauer and Szilard sequences should exactly compensate each other, leading to closed mechanical cycles. On the other hand, if oscillations are too fast, the mechanical work is irreversibly dissipated in the heat bath, therefore damping the mechanical oscillations.

\begin{figure}[t]
\begin{center}
\includegraphics[width=0.5\columnwidth,natwidth=1298,natheight=505]{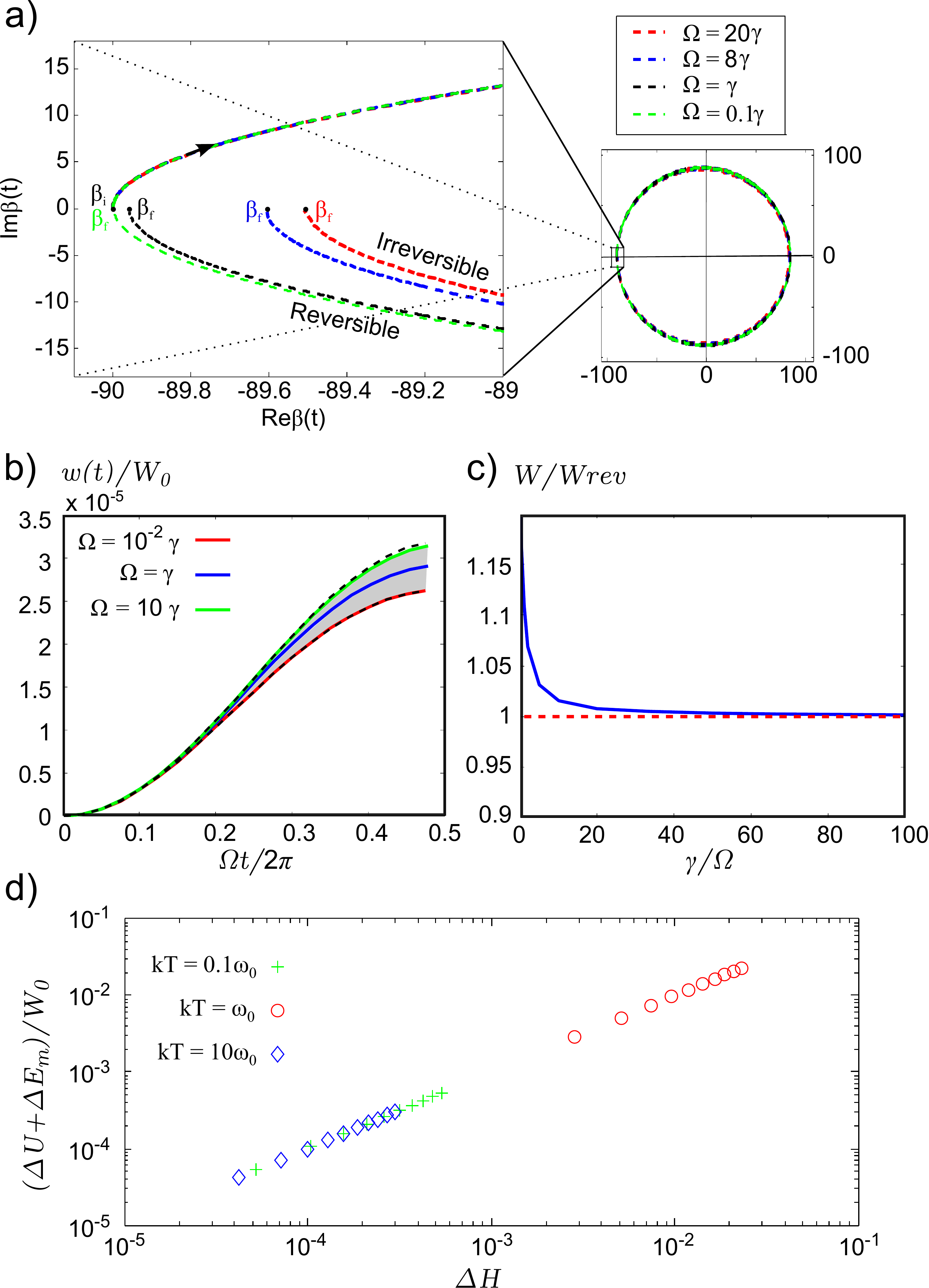}
\end{center}
\caption{a) Right: evolution of $\beta(t)$ over one mechanical oscillation for different values of the mechanical frequency $\Omega$. All evolutions have the same starting point $\beta_i$. Left: zoom around $\beta(t=1/\Omega)$. b) Average work $w$ (in unit of $W_0$) performed on the TLS after time $t$  for different frequencies $\Omega$. The grey area is bounded on the bottom by the reversible work realized when the TLS remains in its steady-state, and on top by the quench work $w_q(t) = P_e(0)(E(t)-E(0))$. c)Work after half a period $W$ divided by the reversible work $W_{rev}$ corresponding to the same transformation performed adiabatically as a function of $\gamma/\Omega$. {\it Simulation parameters} : $\nu_0/\gamma = 10^4$, $g_m/\gamma = 0.1$, ${\color{blue}}\nu_0 / kT = 10$, $\beta(0) = 10^3$.
d) Heat $Q$ deduced from the mechanical energy variation, as a function of the bit  Shannon entropy variation $\Delta H$, for different initial $\beta(0)$ between $10^2$ and $10^3$ and for three different temperatures. The mechanical frequency is $\Omega/\gamma=10^{-3}$, so that the transformations are reversible, and then the two quantities are proportional. Note that with realistic parameters, $\Delta H$ remains smaller than $1$. {\it Simulation parameters} : $g_m/\gamma = 20$, $\nu_0 / \gamma = 5\times 10^3$}
\label{cyril}\end{figure}

First, we have numerically solved equations \eqref{eq:Bloch-beta} to \eqref{eq:Bloch-s}, and derived the mean evolution of the MO complex state $\beta(t)$ using a set of realistic parameters. Results are plotted in fig.\ref{cyril}a for different mechanical frequencies $\Omega$, starting from the same initial state $\beta(t=0)=\beta_i$. We remind that $\delta(t) = g_m \mathrm{Tr}_m[\rho_m(t) (b+b^\dagger)] = 2g_m \Re(\beta(t))$, where $\Re$ stands for the real part.  Work induced modulations of the mechanical energy result in an ovoid shape of the cycle. In practice, the difference between the minimal and maximal deflection that is related to work extraction can be accurately measured after sufficiently long integration time, allowing to get rid of quantum fluctuations and Brownian motion \cite{Sanii}. In the left panel, we have zoomed on $\beta(t)$ after one mechanical oscillation, around $t=1/\Omega$. As the figure shows, the quasi-static regime allows reaching reversibility in the thermodynamic sense, and results in closed mechanical cycles. For higher values of $\Omega$, the quasi-staticity condition is broken. Therefore, extra-work is dissipated in the bath, such that the radius $|\beta(t)|$ of the trajectory displays a net decrease after one period. In Fig.\ref{cyril}b and c, we have represented the work $w(t)$ performed on the TLS during the erasure step. Reversibility is reached as soon as $\Omega \leq 10^{-2} \gamma$, which is easily attained in opto-mechanical devices where $\Omega$ typically ranges from a few kHz to a few MHz, while $\gamma$ scales like a few GHz.

Restricting our study to the quasi-static case, we have checked that Clausius equality could quantitatively be established. In the case of isothermal transformations under study here, it reads

\begin{equation}
- Q = W_0 \Delta H = -\Delta U + W \label{PPT}
\end{equation}

where $\Delta U$ is the change of the TLS internal energy, and $\Delta H$ the TLS Shannon entropy variation. Practically, both quantities can simply be measured by performing fast QND readout of the qubit state at the time where minimal and maximal elongations are reached.
Formula (\ref{PPT}), stating that heat exchanges are proportional to entropy variations of the TLS, is valid even in the case of partial information-to-energy conversions where $|\Delta H| = |H_f - H_i|<1$. This is especially interesting in the case of opto-mechanical systems, where the elastic limit imposes that $\Delta \nu \ll \nu_0$, such that $\Delta E\ll E$.  Therefore, entropy variations are typically bounded to $|\Delta H| \sim 0.1$, which is independent of the parameters used and is only due to the physics of the TLS-mechanical coupling. Fig.\ref{cyril}d shows the average heat $Q$ deduced from the mechanical energy variation from the formula $Q = \Delta U+\Delta E_m$. $Q$ is plotted as a function of the TLS Shannon entropy variations $\Delta H$, for different temperatures of the bath. The expected proportionality between the two quantities clearly appears in the figure : this is a major signature of reversibility, which manifests the proportionality between energy and information, and lies at the basis of most protocols of quantum information thermodynamics. In this case, reversibility is attained in a battery of finite energy, where the back-action from the TLS is visible. Our results offer a new striking illustration that the classical behavior of a physical system is less related to its size, than to the fact that it does not get entangled with another system \cite{Patrice, luiz}. As both conditions of reversibility and finite size battery are met, this proposal paves the way towards direct evidences of optimal information-to-energy conversion in the battery itself.

\section{Conclusion}
We have shown that a hybrid opto-mechanical system is a promising candidate to investigate thermodynamics of information at the single bit level. This device provides direct experimental access to elementary work exchanges between a battery and a single bit coupled to a heat bath. Our study shows that direct monitoring of reversible information-to-energy conversions inside the battery is conceivable in the near future. The TLS embedded in hybrid systems are also good quantum bits. Consequently, the study of heat engines based on one or a few entangled qubits is a natural follow-up of the present work. Hybrid systems could therefore offer an experimental playground to investigate the peculiarities of quantum information, like for instance the work extractable from entanglement \cite{DelRio,Oppenheim, Zurek}, or from engineered heat baths \cite{Poyatos}.

\begin{acknowledgments}
The authors thank O. Arcizet, G. Bulnes Cuetara, M. Clusel, B. Huard, E. Mascarenhas, G. Milburn, J.M. Raimond, M. Santos, V. Vedral and P. Verlot for fruitful exchanges. M.R. acknowledges the ERC starting grant "Handy-Q" n°258608. A.A. acknowledges the support of the ANR-13-JCJC "INCAL".
\end{acknowledgments}

\end{document}